\documentclass[conference]{IEEEtran}
\IEEEoverridecommandlockouts

\usepackage[vlined,linesnumbered,noresetcount]{algorithm2e}
\usepackage{enumitem}
\usepackage{xspace}
\usepackage{float}

\usepackage{pgfplots,pgfplotstable}
\usetikzlibrary{patterns}
\pgfplotsset{compat=1.14}

\usepackage{xcolor}
\usepackage{hyperref}

\begin{document}

\SetKwBlock{SubAlgoBlock}{}{end}
\newcommand{\SubAlgo}[2]{#1 \SubAlgoBlock{#2}}
\let\oldnl\nl
\newcommand{\nonl}{\renewcommand{\nl}{\let\nl\oldnl}}

\newtheorem{question}{Question}

\SetKw{WhenReceived}{when received}
\SetKw{KwTo}{to}
\SetKw{Send}{send}
\SetKw{KwFrom}{from}
\SetKw{Fun}{function}

\newcommand{\System}{PBFT$\star$\xspace}

\newcommand{\REQUEST}{{\tt REQUEST}}
\newcommand{\PREPREPARE}{{\tt PRE\_PREPARE}}
\newcommand{\PREPARE}{{\tt PREPARE}}
\newcommand{\COMMIT}{{\tt COMMIT}}
\newcommand{\REPLY}{{\tt REPLY}}
\newcommand{\CHECKPOINT}{{\tt CHECKPOINT}}
\newcommand{\PROBE}{{\tt PROBE}}
\newcommand{\PROBEACK}{{\tt PROBE\_ACK}}
\newcommand{\NEWCONFIG}{{\tt NEW\_CONFIG}}
\newcommand{\REPLICATELOGS}{{\tt SYNC\_STATE}}
\newcommand{\VIEWCHANGE}{{\tt VIEW\_CHANGE}}
\newcommand{\NEWVIEW}{{\tt NEW\_VIEW}}

\newcommand{\correct}{\mathcal{C}}
\newcommand{\status}{{\sf status}}
\newcommand{\reconstatus}{{\sf rec\_status}}
\newcommand{\probing}{{\sf probing}}
\newcommand{\reconfiguring}{{\sf reconfiguring}}
\newcommand{\txn}{{\sf txn}}
\newcommand{\nextv}{{\sf next}}
\newcommand{\phase}{{\sf phase}}
\newcommand{\payload}{{\sf payload}}
\newcommand{\decision}{{\sf dec}}
\newcommand{\vote}{{\sf vote}}
\newcommand{\ballot}{{\sf view}}
\newcommand{\cballot}{{\sf new\_view}}
\newcommand{\aballot}{{\sf v}}
\newcommand{\vvepoch}{\mathit{view}}
\newcommand{\vaballot}{\mathit{cballot}}
\newcommand{\vcballot}{\mathit{cur\_view}}
\newcommand{\vballot}{\mathit{v}}
\newcommand{\vinit}{\mathit{init}}
\newcommand{\vphase}{\mathit{phase}}
\newcommand{\vpayload}{\mathit{payload}}
\newcommand{\vtxn}{\mathit{txn}}
\newcommand{\vdecision}{\mathit{dec}}
\newcommand{\vvote}{\mathit{vote}}
\newcommand{\dest}{{\sf dest}}
\newcommand{\shards}{{\sf shards}}
\newcommand{\filter}{{\sf filter}}
\newcommand{\proc}{{\sf proc}}
\newcommand{\client}{{\sf client}}
\newcommand{\coord}{{\sf coord}}
\newcommand{\noop}{{\tt nop}}
\newcommand{\origin}{{\sf origin}}
\newcommand{\leader}{{\sf leader}}
\newcommand{\pos}{\mathit{pos}}
\newcommand{\pload}{\mathit{pload}}
\newcommand{\cert}{{\sf cert}}
\newcommand{\init}{{\sf init}}
\newcommand{\vcert}{\mathit{cert}}
\newcommand{\valid}{{\sf valid}}
\newcommand{\ValidNewLeader}{{\sf ValidNewLeader}}
\newcommand{\ValidNewView}{{\sf ValidNewState}}
\newcommand{\quorum}{{\sf quorum}}
\newcommand{\accepted}{{\sf accepted}}
\newcommand{\decided}{{\sf decided}}
\newcommand{\dep}{{\sf prem}}
\newcommand{\length}{{\sf length}}
\newcommand{\vdep}{\mathit{prem}}
\newcommand{\pr}{P}
\newcommand{\M}{{\sf members}}
\newcommand{\Mlast}{{\sf Mlast}}
\newcommand{\blast}{{\sf blast}}
\newcommand{\Servers}{{\sf S}}
\newcommand{\vmembers}{\mathit{members}}
\newcommand{\vleader}{\mathit{leader}}
\newcommand{\vm}{\mathit{M}}
\newcommand{\vl}{\mathit{p}}
\newcommand{\zk}{\textbf{CS}}
\newcommand{\recovery}{{\sf initialized}}
\newcommand{\vrecovery}{\mathit{initialized}}
\newcommand{\pballot}{{\sf probed\_view}}
\newcommand{\uballot}{{\sf upper\_view}}
\newcommand{\rballot}{{\sf recon\_view}}
\newcommand{\lballot}{{\sf last\_view}}
\newcommand{\rshard}{{\sf recon\_shard}}
\newcommand{\pmembers}{{\sf members}}
\newcommand{\rmembers}{{\sf recon\_members}}
\newcommand{\rleaders}{{\sf recon\_leaders}}
\newcommand{\palive}{{\sf palive}}
\newcommand{\Conf}{{\sf C}}
\newcommand{\tid}{\mathit{tid}}
\newcommand{\tshards}{\mathit{tshards}}
\newcommand{\tpayload}{\mathit{payload}}
\newcommand{\tpayloads}{\mathit{payloads}}
\newcommand{\coordpayloads}{{\sf payloads}}
\newcommand{\vpload}{\mathit{pl}}
\newcommand{\fpload}{{\sf pload}}
\newcommand{\fploadprocess}{{\sf topload}}

\newcommand{\lltcsone}{{\sf TCS-LL}}

\newcommand{\prefix}[2]{#1\mathpunct{\downharpoonleft}_{#2}}

\newcommand{\INIT}{\texttt{INIT}}
\newcommand{\LEADER}{\textsc{leader}}
\newcommand{\FOLLOWER}{\textsc{follower}}
\newcommand{\RECOVERING}{\textsc{reconfiguring}}
\newcommand{\LEADERINIT}{\textsc{leader\_init}}
\newcommand{\LEADERSYNC}{\textsc{leader\_sync}}
\newcommand{\FOLLOWERSYNC}{\textsc{follower\_init}}
\newcommand{\FOLLOWERINIT}{\textsc{follower\_init}}
\newcommand{\LEADERPREPARE}{\textsc{leader\_accept}}
\newcommand{\FOLLOWERPREPARE}{\textsc{follower\_accept}}
\newcommand{\LEADERCOMMIT}{\textsc{leader\_commit}}
\newcommand{\START}{\textsc{start}}
\newcommand{\PREACCEPTED}{\textsc{pre-accepted}}
\newcommand{\ACCEPTED}{\textsc{prepared}}
\newcommand{\DECIDED}{\textsc{decided}}
\newcommand{\COMMITTED}{\textsc{committed}}
\newcommand{\ABORT}{\textsc{abort}}
\newcommand{\PREPARED}{\textsc{prepared}}
\newcommand{\PROPOSED}{\textsc{proposed}}
\newcommand{\RECOVERY}{\textsc{recovery}}
\newcommand{\TRUE}{\textsc{true}}
\newcommand{\FALSE}{\textsc{false}}
\newcommand{\READYSTATUS}{\textsc{ready}}
\newcommand{\PROBING}{\textsc{probing}}
\newcommand{\INSTALLING}{\textsc{installing}}

\newcommand{\D}{\mathcal{D}}
\newcommand{\Sh}{\mathcal{S}}
\newcommand{\Proc}{\mathcal{P}}
\newcommand{\Nodes}{\mathcal{N}}
\newcommand{\Txn}{\mathcal{T}}
\newcommand{\Msg}{\mathcal{M}}
\newcommand{\Shard}{\mathcal{S}}
\newcommand{\Ops}{\mathcal{O}}
\newcommand{\Shards}{\mathcal{S}}
\newcommand{\Payload}{\mathcal{L}}
\newcommand{\Certify}{\ensuremath{{\tt certify}}}
\newcommand{\Produce}{\ensuremath{{\tt produce}}}
\newcommand{\Decide}{\ensuremath{{\tt decide}}}
\newcommand{\inv}{\ensuremath{{\sf inv}}}
\newcommand{\resp}{\ensuremath{{\sf resp}}}
\newcommand{\act}{\ensuremath{{\sf act}}}
\newcommand{\ops}{\ensuremath{{\sf ops}}}
\newcommand{\complete}{\ensuremath{{\sf complete}}}
\newcommand{\pending}{\ensuremath{{\sf pending}}}
\newcommand{\aborted}{\ensuremath{{\sf aborted}}}
\newcommand{\committed}{\ensuremath{{\sf committed}}}
\newcommand{\Conc}{\ensuremath{{\sf Conc}}}
\newcommand{\Prec}{\ensuremath{{\sf Prec}}}
\newcommand{\recv}{\ensuremath{{\sf receive}}}
\newcommand{\send}{\ensuremath{{\sf send}}}
\newcommand{\hist}{h}
\newcommand{\cf}{f}
\newcommand{\cg}{g}

\newcommand{\commit}{\ensuremath{{\sf commit}}}
\newcommand{\Read}{\ensuremath{{\sf read}}}
\newcommand{\Write}{\ensuremath{{\sf write}}}
\newcommand{\readset}{\ensuremath{{\sf readset}}}
\newcommand{\writeset}{\ensuremath{{\sf writeset}}}
\newcommand{\version}{\ensuremath{{\sf version}}}
\newcommand{\val}{\ensuremath{{\sf val}}}
\newcommand{\Val}{\ensuremath{{\sf Val}}}
\newcommand{\Obj}{\ensuremath{{\sf Obj}}}
\newcommand{\Ver}{\ensuremath{{\sf Ver}}}
\newcommand{\fetch}{\ensuremath{{\sf fetch}}}
\newcommand{\trans}{\ensuremath{{\sf trans}}}
\newcommand{\dec}{\ensuremath{{\sf dec}}}
\newcommand{\pref}{\ensuremath{{\sf prefix}}}

\newcommand{\tr}[2]{\iflong{}\S#1\else{}\cite[\S#2]{ext}\fi}
\newcommand{\tra}[2]{\iflong{}(\S#1)\else{}\cite[\S#2]{ext}\fi}

\newcommand{\rt}{\prec_{\text{rt}}}
\newcommand{\decrel}{\prec_{\text{dec}}}

\newcommand{\external}{configuration service\xspace}


\newcommand{\CONFIGPREPARE}{{\tt CONFIG\_PREPARE}}
\newcommand{\CONFIGPREPAREACK}{{\tt CONFIG\_PREPARE\_ACK}}
\newcommand{\RDMAOP}{{\tt RDMA\_OP}}
\newcommand{\RDMAACK}{{\tt RDMA\_ACK}}
\newcommand{\openconnections}{{\sf connections}}
\newcommand{\buffer}{{\sf buffer}}
\newcommand{\pop}{{\sf pop}}
\newcommand{\emptyfunction}{{\sf empty}}
\newcommand{\coorddecisions}{{\sf tdecisions}}
\newcommand{\coordpositions}{{\sf tpositions}}
\newcommand{\emptybuffers}{{\bf flush}\xspace}
\newcommand{\openconnection}{{\bf open}\xspace}
\newcommand{\closeconnection}{{\bf close}\xspace}
\newcommand{\closeallconnections}{{\tt close\_all\_connections}}
\newcommand{\sendrdma}{{\bf send-rdma}\xspace}
\newcommand{\deliverrdma}{{\bf deliver-rdma}\xspace}
\newcommand{\ackrdma}{{\bf ack-rdma}\xspace}
\newcommand{\multiclose}{\xspace{\tt multiclose}\xspace}
\newcommand{\multiopen}{\xspace{\tt multiopen}\xspace}
\newcommand{\updatemembership}{\xspace{\tt update\_members}\xspace}

\newcommand{\Order}{\ensuremath{{\tt order}}}
\newcommand{\Digest}{\ensuremath{{\tt digest}}}
\newcommand{\Getdigest}{\ensuremath{{\tt get\_d}}}
\newcommand{\Getseq}{\ensuremath{{\tt get\_n}}}
\newcommand{\Execute}{\ensuremath{{\tt execute}}}
\newcommand{\Clean}{\ensuremath{{\tt clean}}}
\newcommand{\Client}{\ensuremath{{\tt client}}}
\newcommand{\Valid}{\ensuremath{{\tt valid}}}
\newcommand{\Checkack}{\ensuremath{{\tt check}}}
\newcommand{\msg}{{\sf msg}}
\newcommand{\dig}{{\sf digest}}
\newcommand{\PREPREPARED}{\textsc{pre\_prepared}}
\newcommand{\statesf}{{\sf state}}
\newcommand{\result}{{\sf result}}
\newcommand{\checkpoint}{{\sf checkpoint}}
\newcommand{\vstate}{\mathit{state}}
\newcommand{\nonop}{\textsc{nonop}}


\title{Towards Improving the Performance of BFT Consensus For Future Permissioned Blockchains}
\author{\IEEEauthorblockN{Manuel Bravo, Zsolt Istv\'{a}n}
\IEEEauthorblockA{IMDEA Software Institute, Madrid\\
\{manuel.bravo, zsolt.istvan\}@imdea.org}
\and
\IEEEauthorblockN{Man-Kit Sit}
\IEEEauthorblockA{City University of Hong Kong\\
manksit@cityu.edu.hk}
}

\maketitle

\begin{abstract}

Permissioned Blockchains are increasingly considered in enterprise use-cases, many of which do not require geo-distribution, or even disallow it due to legislation. Examples include country-wide networks, such as Alastria, or those deployed using cloud-based platforms such as IBM Blockchain Platform. We expect these blockchains to eventually run in environments with high bandwidth and low latency modern networks, as well as, advanced programmable hardware accelerators in servers. 

Even though there is renewed interest in BFT consensus algorithms with various proposals targeting Permissioned Blockchains, related work does not optimize for fast networks and does not incorporate hardware accelerators -- we make the case that doing so will pay off in the long run.
To this end, we re-implemented the seminal PBFT algorithm in a way that allows us to measure different configurations of the protocol. Through this we explore the benefits of various common optimization strategies and show that the protocol is unlikely to saturate more than 10Gbps networks without relying on specialized hardware-based offloading. We discuss two concrete ways in which the cost of consensus in Permissioned Blockchains could be reduced in high speed networking environments, namely, offloading to SmartNICs and implementing the protocol on standalone FPGAs. 

\end{abstract}

\section{Introduction}

Blockchain is an emerging technology, considered increasingly often beyond the crypto-currency world for business-to-business use-cases. In contrast to public blockchains such as Bitcoin, that are open systems in which anyone can participate, in business-to-business scenarios the membership of the service is tightly controlled and this permits the use of Byzantine fault tolerant (BFT) consensus protocols at the core of the service to establish a total order of transactions, instead of the more expensive Proof-of-Work-based consensus protocols. Such systems are called \emph{permissioned blockchains}~\cite{androulaki2018hyperledger,brown2016corda,CCF}.
In a permissioned blockchain system, often only a subset of the total number of participating nodes run the BFT protocol~\cite{alastria,kwon2014tendermint} and, in general, members have more control over how and where to run the network~\cite{resilientdb2020vldb}.



Driven by opportunities in blockchain, there has been an increased interest in BFT consensus protocols~\cite{yin2019hotstuff,gramoli2017blockchain,stathakopoulou2019mir}.
Interestingly, the deployment model of permissioned blockchains can be very different from permissionless ones likes Bitcoin. While the latter is typically widely distributed, with bandwidth and latency characteristics much like that of the world wide web, in the permissioned blockchain space, there are use cases where nodes are under tighter control (e.g., those in hosted environments on Amazon Managed Blockchain~\cite{amazon-blockchain} or IBM Blockchain Platform~\cite{ibm-blockchain}), perhaps even geographically confined (e.g., emerging country-wide networks, such as Alastria~\cite{ruiz2020public} in Spain). Nodes of such deployments have access to more bandwidth, lower latency communication than what we associate today with blockchains. Given the increasing presence of programmable hardware devices in public clouds~\cite{firestone2018azure,bosshart2014p4,jin2018netchain}, it is likely that nodes could even rely on these for increasing performance. If successful, blockchain technology will likely replace several database solutions in the area of banking, trading and e-commerce but the performance of today's permissioned blockchains will not be satisfactory. For this reason, it is important to start investigating strategies for increasing the speed of BFT consensus using modern hardware available in the clouds and datacenters. As we show in Section~\ref{sec:motivation}, current BFT consensus implementations are unable saturate bandwidths of 10Gbps and higher, while retaining low latency. 

Our goal is to investigate how far can software get us and to what extent will it be useful, or even necessary, to use hardware accelerators in the future. We apply a experiment-driven approach to quantify the benefits of various existing optimization strategies. For this, we build a framework that integrates 
a streamlined variant of the seminal PBFT~\cite{castro1999practical,castro2002practical} consensus protocol and can be configured at multiple levels. 
Our study reveals that, even after applying various optimizations, achieving 10Gbps performance in software without relying on very large batches is still unlikely -- the road the 100Gbps rates will hence have to involve some forms of hardware accelerators. We can also confirm that contrary to anecdotal evidence, hardware accelerators for cryptographic operations alone will not result in significantly better performance because the biggest cost, even in modestly sized consensus groups, is that of packet parsing and hashing, that is, data-movement related operations.

To alleviate the cost of these operations, we propose two strategies for incorporating specialized hardware, namely, emerging Smart Network Interface cards (SmartNICs) and standalone FPGA nodes to provide line-rate, predictable behavior. We present micro-benchmarks motivating these strategies and discuss their main benefits and open challenges.

\smallskip
\noindent
Overall, this work brings three contributions:
\begin{itemize}

\item We identify the future need for low latency and high bandwidth BFT consensus. Even though today the challenges of Permissioned Blockchains lie in determining the right governance model and integration with data management solutions, if successful, these blockchains will have to provide high throughput, low latency, and the ability to scale with faster networks and more powerful hardware.

\item We provide an open-source framework\footnote{Link removed for double blind submission.} for experimenting with various software and hardware strategies for accelerating PBFT and similar protocols. 

\item Based on measurements carried out with our framework, we identify two specific hardware acceleration scenarios that will be able to saturate 10Gbps and faster networks even for small consensus groups.
\end{itemize}

\section{Motivation}
\label{sec:motivation}

In this section, we motivate the need for investigating performance-related aspects of BFT consensus protocols in environments with high network bandwidth and low latency. We first discuss three use-cases in which consensus nodes are geographically confined by design or have access to high bandwidth networking and their location can be controlled. Second, we show that state-of-the-art BFT consensus cannot efficiently take advantage of fast networks, further motivating the ``de-construction'' of the underlying protocol for measurement purposes. 

\subsection{Use-cases}

\noindent\textbf{Single and Multi-Cloud Hosted Blockchains.}
Several cloud providers are offering hosted blockchain solutions both as Software as a Service (e.g., Azure Blockchain Service), and by simplifying the deployment of open-source, commonly used, networks such as Hyperledger Fabric (e.g., in IBM Cloud Blockchain Platform). Given that a large portion of web-facing applications already run in hosted environments, running blockchains as a ``backend'' in the cloud is a natural step in many scenarios. In cloud environments the blockchain nodes have access to high bandwidth networking and low latencies within regions, and it is also to be expected that with the emergence of more enterprise use-cases for permissioned ledgers, multi-cloud deployments will rely on dedicated links for higher bandwidth across clouds and reduced data movement costs, as it is already the case for CDNs. For this reason, when preparing the next generation of Blockchain-focused BFT consensus libraries, it is crucial to design with high bandwidth (10Gbps and above) and low latency networks in mind.


\medbreak
\noindent\textbf{Regional Replication by Design.}
There are emerging country-wide permissioned blockchain networks such as Alastria~\cite{alastria,ruiz2020public} in Spain, that set out to provide a mechanism for any company within a consortium to interact with any other one through smart contracts that are recognized under the local legislation. For this reason, these kinds of networks are run in geographically more confined environments. Furthermore, in this type of networks, only a small subset of the consortium nodes take part actively in consensus (i.e., running the core operations of the network), making them the critical performance point of the network. Thus, due to these two 
characteristics, we expect the consortium to optimize the environment in which consensus nodes are deployed. Hence,  when this type of networks matures, we expect the group of consensus nodes to run in a environment with high bandwidth and low latency, as well as, have advanced programmable hardware accelerators (already commonly found in servers) at their disposal.

\medbreak
\noindent\textbf{Geo-replicated Systems with Local Optimizations.}
The recent work by Gupta et al.~\cite{resilientdb2020vldb}, called ResilientDB, argues that, in order to build practical geo-distributed databases based on blockchain technology, it is crucial to minimize cross-region communication between nodes without reducing reliability or availability guarantees. The proposed design has a hierarchical consensus mechanism that runs several BFT consensus groups, with nodes close by, and performs a geo-replication using a step that requires only a linear number of messages in the failure-free case. Thus, given that the performance of the consensus groups would drive the overall performance of the system and that nodes within a group are close by, one can expect these groups to be deployed in a environment with high network resources.


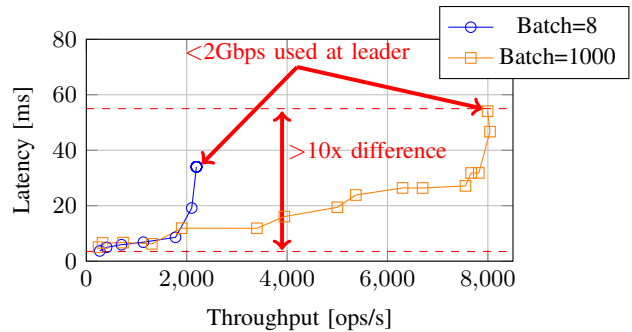
\begin{figure}[pt]
  \centering
\begin{tikzpicture}  
  \pgfplotstableread{ 
T8 L8 T1000 L1000
267	3.689	244	5.036
404	4.981	323	6.570
701	5.946	731	6.700
1136	6.843	1306	6.254
1777	8.599	1900	11.898
2100	19.170	3399	11.853
2194	34.019	3945	16.089
2194	34.019	5001	19.429
2194	34.019	5370	23.854
2194	34.019	6300	26.408
2194	34.019	6700	26.388
2194	34.019	7553	27.140
2194	34.019	7666	31.800
2194	34.019	7819	31.816
2194	34.019	8042	46.710
2194	34.019	7994	54.091
  }\datatable
  \begin{axis}[
   legend style={at={(1.05,1.15)},
    anchor=north},
  legend columns=1,
  font=\small,
  width=0.4\textwidth,
  height=0.25\textwidth,
  ylabel={Latency [ms]},
  xlabel={Throughput [ops/s]}, 
  ymin=0, 
  ymax=80,
  xmin=0,
  xmax=8500,
  grid=major,
  ]

  \node[color=red] at (axis cs: 4200,75) {$<$2Gbps used at leader};
  \draw[->, line width=1.5pt, color=red] (axis cs:4200,70) -- (axis cs:2300,35);
  \draw[->, line width=1.5pt, color=red] (axis cs:4200,70) -- (axis cs:7900,55);

  \draw[<->, color=red, line width=1.5pt] (axis cs:3900,4) -- (axis cs:3900,54);
  \draw[color=red, dashed] (axis cs:0,55) -- (axis cs:8500,55);
  \draw[color=red, dashed] (axis cs:0,3.5) -- (axis cs:8500,3.5);
  \node[color=red] at (axis cs: 5600,40) {$>$10x difference};

  \addplot [blue, mark=o] table [x=T8, y=L8] {\datatable};  
  \addplot [orange, mark=square] table [x=T1000, y=L1000] {\datatable};

  \legend{Batch=8, Batch=1000}
  \end{axis}
  \end{tikzpicture}  
  \caption{\label{fig:bft-smart}State-of-the-art BFT consensus libraries fail to efficiently utilize 10Gbps LAN connections (BFT-Smart with 7 nodes, 4KB payload, no logging)}
\end{figure}

\begin{figure*}[h!]
  \centering
  \includegraphics[width=0.8\linewidth]{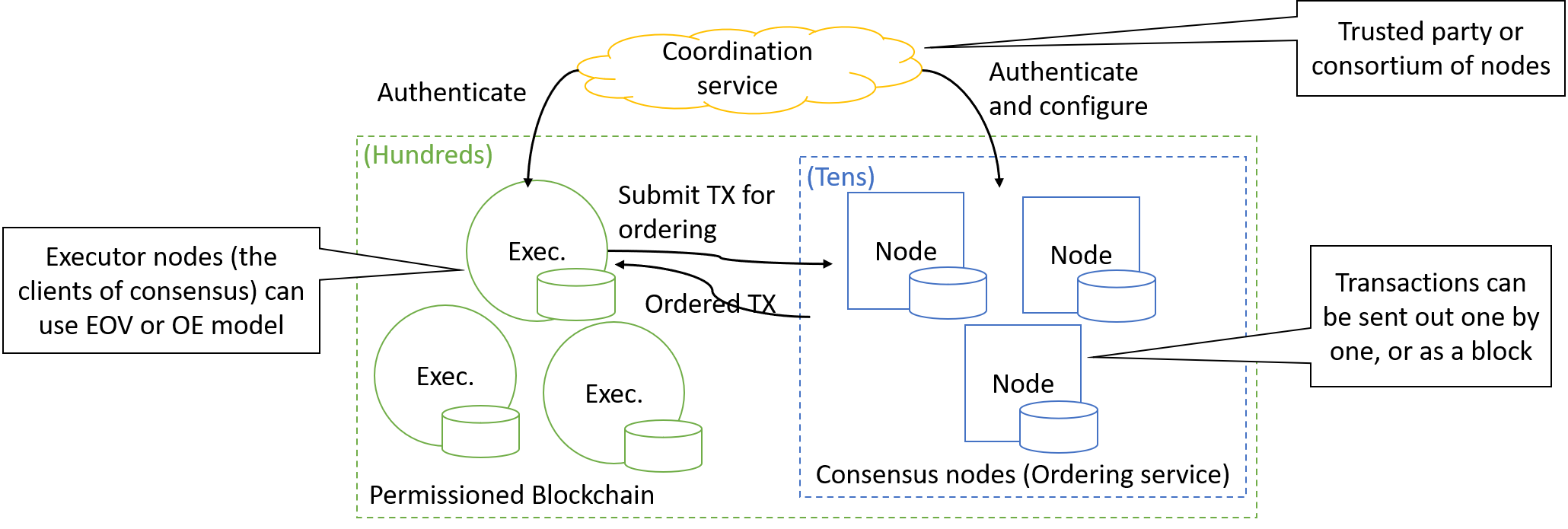}
  \caption{\label{fig:sys-assumptions} In permissioned blockchains a coordination service authenticates nodes in the network and performs configuration. There are significantly less consensus nodes than regular member nodes and their churn is minimal.}
\end{figure*}

\subsection{State-of-the-art BFT Consensus and Fast Networks}
To experimentally show that state-of-the-art BFT consensus protocols are unable to take advantage of fast networks, we measure the performance of BFT-Smart~\cite{bessani2014state}, one of the most optimized BFT consensus libraries available, on such networks. We have chosen BFT-Smart for a few reasons: (i) with more that five years of development, we consider it a serious attempt of implementing a highly-performant BFT consensus, (ii) it integrates a consensus protocol similar to the seminal PBFT protocol, on which most of the modern BFT consensus protocols are based, (iii) it includes a set of optimizations and refinements aiming to enhance performance such as multi-core awareness and the use of \emph{cheaper} cryptographic operations when possible, and (iv) it is open source and actively supported.

We configured BFT-Smart with 7 nodes with RSA1024 signatures and MACs among the nodes, on server-grade machines connected over a 10Gbps LAN (see the Experiments section for details). We experiment with two batch sizes: 1000 request per batch (the default), and a smaller one with 8 requests per batch. 
Figure~\ref{fig:bft-smart} reports the average latency vs. throughput achieved by BFT Smart when running the YCSB benchmark\footnote{Default YCSB configuration from the BFT-Smart repository: updates only and a single field per entry to avoid computational overhead in the nodes} with 4KB values. The experiments show that the leader node is far from saturating the network connection. In fact, it uses significantly less than 2Gbps-worth of bandwidth even at saturation -- showing that there is a need to explore how to design BFT consensus solutions that can saturate 10Gbps bandwidth, and beyond.

Furthermore, BFT-Smart is unable to keep latencies consistently low while delivering high throughput. As Figure~\ref{fig:bft-smart} shows, BFT-Smart exhibits a latency that is more than an order of magnitude greater than the network's response time. This is mainly because, in order to enhance throughput, BFT-Smart employs batching aggressively: it composes large batches of messages in an attempt to reduce the overhead of consensus. Figure~\ref{fig:bft-smart} shows that when significantly reducing the batch size, the difference in throughput is stark: the throughput drops by 4x. In this work, we investigate the inherent tension between latency and throughput in BFT consensus protocols.

\section{Background}

\subsection{Permissioned Blockchains}

Public blockchains are often associated with crypto-currencies and are characterized by the fact that nodes can join without permission. For this reason, many of these blockchains implement Proof of Work, or similar, consensus methods~\cite{vukolic2015quest} and are designed without assumptions about the nodes. Permissioned ledgers~\cite{androulaki2018hyperledger,brown2016corda,CCF}, in contrast, rely on a trusted service or consortium to authenticate nodes when joining the blockchain, but do not assume trustworthiness of nodes. This is useful in business-to-business scenarios where the goal of the blockchain is not to offer anonymity but rather to logically centralize data and run tamper-free ``smart contracts'', application logic, on it. Example use-cases include ones in health care~\cite{azaria2016medrec}, supply chain management~\cite{korpela2017digital}, etc., but also in areas such as banking and capital markets~\cite{buehler2015beyond}. In these scenarios all actors in the system are known but they want to protect against malicious actions from the others. 

As seen in Figure~\ref{fig:sys-assumptions}, permissioned blockchains are composed of executor nodes and consensus nodes. For generality, we consider these sets of nodes to be disjoint but, in practice, a node could implement both roles. Clients of the blockchain system, i.e., users, are external to this illustration. The coordination service shown in the image is the trusted third party or consortium of nodes (that all members of the blockchain trust) that authenticates nodes, configures the network, etc. 

Permissioned ledgers usually provide one of two execution models: order-execute (OE), or execute-order-validate (EOV). The first means that smart contracts with their specific inputs are submitted first to the ordering service, that is, the consensus nodes, and then executed on all nodes of the network. The EOV model simulates the contract execution on a subset of the nodes and submits the resulting ``read-write set'' for ordering. The nodes receive these from the ordering service and update their state if the read-write sets do not conflict with the ledger state. Even though these two models offer different trade-offs, from the perspective of the underlying consensus logic, they are very similar. For this reason in this work we investigate BFT ordering without assuming one or the other execution model.

\begin{figure*}[t]
  \centering

  \includegraphics[width=0.9\linewidth]{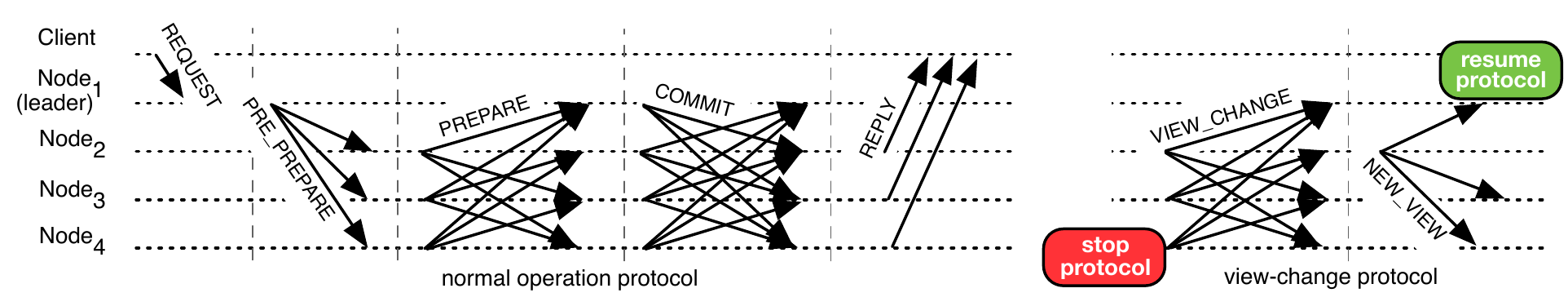}
  \caption{\label{fig:pbft} PBFT communication pattern during failure-free operations and recovery.}
\end{figure*}

The question of how executor nodes ``get'' the ordered transactions is also orthogonal to our investigation. We make the assumption that, in general, executor nodes are interested in pulling transactions from the ordering service as soon as they are ordered (they can access at the block granularity to recover and to gossip). With this assumption it is beneficial to explore not only the throughput but also the latency improvements one could add to BFT protocols. This is relevant as there is recent work on exploring how the throughput~\cite{gorenflo2019fastfabric,sharma2019blurring,amiri2019caper} and latency~\cite{istvan2018streamchain} of blockchains can be improved substantially in the presence of fast networks. 

\subsection{PBFT}


The seminal PBFT~\cite{castro1999practical} protocol is one of the most well-studied protocols and that is why we use it in our study. The protocol requires a minimum of $3f+1$ nodes to tolerate $f$ faulty nodes. For simplicity, in this section, we consider the variant of PBFT that uses public-key signatures. We depict its communication pattern in Figure~\ref{fig:pbft}.

The protocol proceeds in rounds called \emph{views}. On each view, one node is the leader and the rest are its followers. The protocol moves to the next view only if the leader is faulty or if asynchrony prevents the protocol to make progress. The process of changing view is called \emph{view-change}. At a given view $v$,
the leader sequences and proposes client's
request in a $\PREPREPARE$ message to the followers. When a follower receives a $\PREPREPARE$ message, it first validates the leader's proposal by checking the authenticity of the client's request and that it does not have another client request already assigned to that sequence number. If the follower \emph{accepts} the request, it sends a $\PREPARE$ message to all nodes. When a node (leader or follower) receives $2f+1$ matching $\PREPARE$ messages, it considers the request as \emph{prepared} and sends a $\COMMIT$ message to all nodes. Intuitively, theses first two steps of the protocol (leader's proposal and the all-to-all communication step) ensure that correct nodes agree on a total order
for the requests within a view. When a node receives $2f+1$ matching $\COMMIT$ messages for a client's request and all requests with a lower sequence number have been committed locally, the node considers the request as \emph{committed} and replies to the client. This second all-to-all communication step, together with the view-change protocol, guarantees that correct replicas agree on the sequence numbers assigned to committed requests even when committing them across views.
Finally, a client waits for $f+1$ matching replies before accepting the result.

When a node wants to move to the next view $v+1$, it first stops executing the protocol and sends a $\VIEWCHANGE$ message to all nodes. A node sends in its $\VIEWCHANGE$ message all client requests that could have been committed. When the leader of $v+1$ gathers $2f+1$ of these messages, it computes the final set of potentially committed requests $\mathcal{O}$ and sends it in a $\NEWVIEW$ message to its followers, together with the $2f+1$ $\VIEWCHANGE$ messages based on which $\mathcal{O}$ was computed. Upon reception of a $\NEWVIEW$ message, a follower first verifies the correctness of $\mathcal{O}$. Then, it adds the new information to its local state and resumes execution.

\section{Experimental Framework}
\label{sec:implementation}

Our main goal is to study the effect of various optimizations of BFT consensus with the permissioned blockchain use-case in mind. Our framework integrates a streamlined variant of PBFT~\cite{castro1999practical,castro2002practical}. We expect that the findings of this work are directly applicable to other BFT consensus protocols~\cite{gueta2019sbft, kotla2007zyzzyva, yin2019hotstuff, stathakopoulou2019mir}, that mostly are optimized variants of PBFT.

One difference between our experimental framework and typical BFT consensus implementations is that we do not rely on batching by default. Our goal is to study the cost of running consensus without batching, or with very small batches, to ensure that the latency of the protocol is representative of the underlying network latency. 

\begin{figure}[t]
  \centering
  \includegraphics[width=0.99\linewidth]{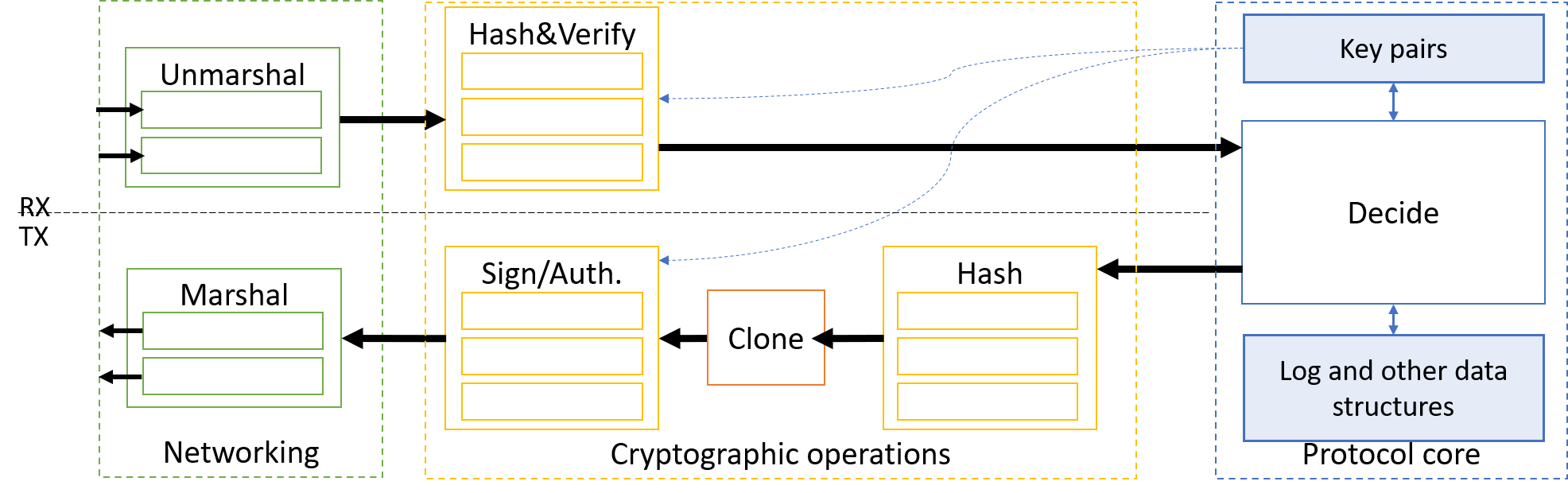}  
  \caption{\label{fig:sw-arch} The consensus logic is implemented as a software pipeline, with data parallel execution for more compute-intensive steps.}
\end{figure}

\subsection{Design}

Implementing consensus protocols on multi-core CPUs leads to the question of how to exploit the available parallelism given that at their core, all protocols, including PFBT, require a serial decision making step. Many BFT and CFT consensus implementations adopted a pipelined architecture~\cite{clement2009upright,santos2013achieving,bessani2014state} and in this work we do the same. At its core, our framework integrates \System, a streamlined variant of PBFT.

Pipelined architectures are also beneficial because 
it is easier to envision the integration of various accelerators than into monolithic ones. One drawback, however, of the pipelining approach is that performance can be bottlenecked on the slowest pipeline stage; for this reason we will consider parallelism both across pipeline stages and within pipeline stages wherever possible.

The framework is parametric to the type of cryptographic operation used for the authentication of different message types: one can choose between public-key (PK) signatures or message authentication codes (MACs). We use in our experimental analysis the following three configurations:
\begin{enumerate} 
\item
\emph{Off-the-shelf:} This variant makes no assumptions about the system around it and uses PK signatures on all messages.

\item
\emph{Algorithmic optimizations:} It replaces PK signatures with MACs on all inter-node messages, similar to the optimizations described in the journal version of PBFT~\cite{castro2002practical}.

\item
\emph{Domain-optimized for Permissioned Blockchains:} It further eliminates PK signatures on responses to clients because client requests (transactions to order) will be logically packed into blocks and it is enough to sign the blocks with a PK and responses to clients with MACs.
\end{enumerate}

Incoming messages from clients are by default using private-key signatures in all variants to counter malicious clients (i.e., big-MAC attack~\cite{clement2008bft}). 

Furthermore, one can finely tune the batching
sizes to explore the latency-throughput tradeoff
when combined with cryptographic operations; and the parallelism of tasks, such as the computation or
verification of cryptographic signatures.

Figure~\ref{fig:sw-arch} shows the stages in our implementation that follow the main steps of the protocol. The only part of the behavior that requires more explanation is the sending hashing and signing: The \emph{Hash(TX)} step hashes messages to prepare them for signing/authentication. The operation in carried out in parallel for multiple messages (round-robin). Once messages have their hashes computed, they are \emph{cloned} to create multiple copies of them to be sent later to individual recipients. If, for instance, a $\PREPREPARE$ message has to be sent to all participants, it is hashed once in the previous step and then cloned in this step for each recipient. The \emph{Sign/Auth.} step computes the signature/MAC to be attached to each message in parallel. This layout is advantageous because by default our implementation is set up to compute different signatures and MACs for each recipient. For protocol variants in the evaluation that only use public-key signatures, this step is merged with hashing to avoid redundant computation.

\subsection{Implementation}
\label{sec:implementation}

\medskip
\noindent
\emph{Pipeline Execution Model.}
The \System's pipeline decouples the building blocks of the protocol and allows for future exploration of different acceleration opportunities (Figure~\ref{fig:sw-arch}). The stages are as follows:
\begin{enumerate} 

\item \emph{Unmarshal}: incoming messages are received on TCP/IP connections and unmarshaled, using one thread for each individual node and client.

\item \emph{Hash (RX) and Verify}: each message is signed by its sender using their private-key or authenticated using a MAC. In either case, the message contents need to be hashed and this hash has to be compared to the one in the signature/auth. This operation is performed by multiple threads in a data-parallel manner using round-robin scheduling to maintain FIFO order of messages.

\item \emph{Decision}: This is where the protocol itself is running. Depending on the internal state and the content of the incoming message, this step will produce one or multiple messages with a list of recipients each.

\item \emph{Hash (TX)}: This step hashes messages to prepare them for signing/authentication. The operation in carried out in parallel for multiple messages (round-robin).

\item \emph{Clone}: Once messages have their hashes computed, this step creates multiple copies of them to be sent later to individual recipients. If, for instance, a $\PREPREPARE$ message has to be sent to all participants, it is hashed once in the previous step and then cloned in this step for each recipient.

\item \emph{Sign/Auth}: This step computes the signature/MAC to be attached to each message in parallel. This layout is advantageous because by default \System is set up to compute different signatures and MACs for each recipient. For protocol variants in the evaluation that only use public-key signatures, this step is merged with hashing to avoid redundant computation.

\item \emph{Marshal}: Signed messages with one recipient each are enqueued on threads representing individual TCP/IP sockets. These perform the serialization of the messages.
\end{enumerate}

\medskip
\noindent
\emph{Implementation Decisions and Optimizations.}
We implemented our prototype in Golang relying heavily on goroutines for parallelism. We use the SHA256 cryptographic hash function to compute digests, RSA-2048 for signatures and AES with 256bit keys for MACs, using default Golang libraries. The messages exchanged between nodes are serialized using Protocol Buffers and follow a similar layout with a fixed set of integer fields followed by a variable length ``attached data'' field.

Since we do not want to restrict the applicability of our prototype, the messages coming from the clients are treated as BLOBs that are recorded in a log. They are not applied, in the traditional sense, to a state database. This is because in most blockchain systems the ordering service does not actually look at the contents of ``transactions''. And even if some processing of these transactions would be necessary, it can be performed off the critical path. This choice, however, introduces a question related to state compaction. While in state machine replication this can happen implicitly at specific intervals on all nodes (e.g., after each successful checkpoint), in a scenario we are looking at, compaction can only be done from the ``outside'' when all clients can agree. In our prototype we keep a log of 10k operations in memory that acts as a circular buffer and we have set the checkpointing frequency to 500 messages to keep most of the data structures fit in cache. In a full implementation, the log would have to be written to disk asynchronously for durability, and a suitable compaction method would have to be chosen.

Similarly, there is a decision to be had in the system whether the blockchain entries can be gossiped or not by the clients. If no (our default assumption), it is sufficient to use MACs to authenticate messages between ordering nodes and clients and for each new client to read blocks directly from the ordering service when recovering state. If yes, the nodes need to sign client responses with a PK signature. Our prototype offers both options and the signatures can be enabled with an environment variable. Incoming messages from clients are by default using private-key signatures, even if the answers are only signed with MACs, to counter malicious clients (big-MAC attack~\cite{clement2008bft}). 

\section{Study of Protocol Variants}
\label{sec:study}

All experiments are performed on a consensus group of 15 nodes, with the clients issuing either requests with 512\,B or 4096\,B values. We chose these two sizes close to the average size of a Bitcoin transaction~\cite{bitcoin-tx-size} and more general smart contracts reported in Fabric~\cite{androulaki2018hyperledger}. We do not use batching, unless otherwise stated, because with new latency-focused designs for permissioned blockchains~\cite{istvan2018streamchain,gorenflo2019fastfabric,amiri2019caper} and the increasingly fast networking in cloud environments, we believe it is important to investigate the protocol without compromising its latency. The theoretical maximum throughput, excluding TCP/IP overhead and without batching, over 10Gbps network is just above 80\,kops/s for 512\,B and 19\,kops/s for 4096\,B values.
We perform our evaluation on a 10\,Gbps cluster of 24 machines with 6 core Intel Xeon E-2186G CPUs. All machines run recent versions of Debian linux and Go v1.10. 




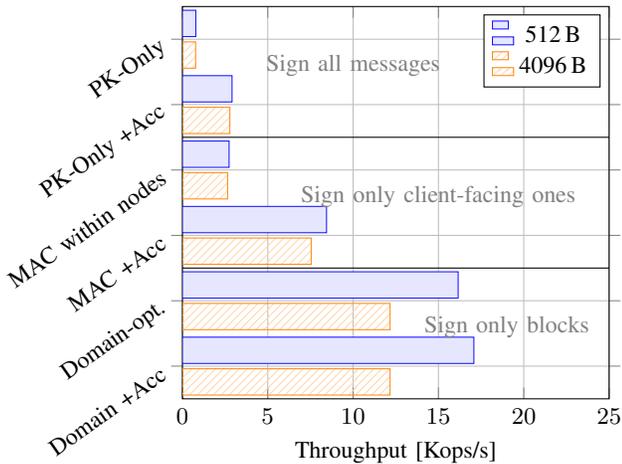
\begin{figure}[pt]
\begin{tikzpicture}
\pgfplotstableread{ 
  Label   Bandwidth 
  {Domain +Acc} 17.076  
  {Domain-opt.} 16.155
  {MAC +Acc} 8.443
  {MAC within nodes} 2.733
  {PK-Only +Acc}  2.907
  {PK-Only}      .784
}\datatable
\pgfplotstableread{ 
  Label   Bandwidth 
  {Domain +Acc} 12.164
  {Domain-opt. } 12.164
  {MAC +Acc} 7.545
  {MAC within nodes} 2.648
  {PK-Only +Acc}  2.778
  {PK-Only}      .773 
}\datatablelarge
\begin{axis}[
legend reversed,
font=\small,
width=0.4\textwidth,
height=0.375\textwidth,
xbar,   
xmin=0,         
ytick=data,     
yticklabels from table={\datatable}{Label},  
y tick label style={rotate=35,anchor=east},  
xtick={0,5.000,10.000,15.000,20.000,25.000,30.000,350.00},
xmin=0,
xmax=25.000,
ymin=-0.5,
ymax=5.5,
bar width=0.35cm,
xlabel={Throughput [Kops/s]},
grid=both,
]
\addplot [orange!90, pattern=north east lines, pattern color=orange!30] table [x=Bandwidth, y expr=\coordindex] {\datatablelarge};
\addplot [blue!90, fill=blue!10] table [x=Bandwidth, y expr=\coordindex] {\datatable};

\node[color=gray, rotate=0] at (axis cs: 10.000,4.6) {Sign all messages};
\draw (axis cs:0,3.5) -- (axis cs:35.000,3.5);
\node[color=gray, rotate=0] at (axis cs: 15.000,2.6) {Sign only client-facing ones};
\draw (axis cs:0,1.5) -- (axis cs:35.000,1.5);
\node[color=gray, rotate=-0] at (axis cs: 19.000,0.6) {Sign only blocks};

\legend{4096\,B, 512\,B}
\end{axis}
\end{tikzpicture}

\caption{\label{fig:tput-variants}We emulate different configurations with 15 nodes to quantify the expected benefits of hardware accelerators and protocol optimizations. The results show that the avoidance of PK cryptography is the most important performance factor.}
\end{figure}

\subsection{What is the performance of off-the-shelf protocols on fast networks?}
In Figure~\ref{fig:tput-variants} we show the throughput of ``off-the-shelf'' BFT, using public-key (PK) cryptography for signing all messages. The numbers are low (less than 800\,ops/s) due to the high computational cost of creating signatures, even though the system uses all cores of the machine. In this scenario, any method of accelerating the crypto operations will be beneficial. In our example, using a faster module for cryptographic operations (``+Acc'') leads to 4X increase in throughput.

\subsection{How much can be gained by using MACs instead of signatures?}
The second class of BFT deployments replace PK signatures on inter-node traffic with MACs~\cite{castro2002practical} that are orders of magnitude cheaper to compute than signatures. The throughput of the system increases to more than 2.6\,Kops/s and is, in fact, on par with the PK-Only version with acceleration. Since the nodes still spend significant resources on signing responses to clients with a PK, adding crypto acceleration is beneficial. It brings, however, a smaller benefit than in the previous case, that is, around 2.5X vs. 4X. Even though in this experiment we do not consider batching, it is worth pointing out that the MAC-based version is an upper bound of throughput for the off-the-shelf version with batching at the leader -- however batching hundreds of requests to amortize the PK signing cost on protocol messages would impact latencies significantly.

\subsection{How much can be gained by optimizing to the domain of permissioned blockchains?}
If nodes use MACs between themselves, as well as, to answer to clients, performance increases significantly, even when those rely on acceleration, there is almost a 2X difference in throughput for small values. 
Using the domain-optimized version and issuing large 4096\,B values, it is possible to achieve more than 60\% of the theoretical maximum throughput without relying on any type of batching. For smaller, 512\,B values, only 25\% of the maximum is reached. It has to be noted that for completeness, signatures have to be computed periodically (e.g. at checkpoints) on the data to allow for recovery at a coarser granularity, as well as to allow clients to exchange ``blocks'' among themselves.

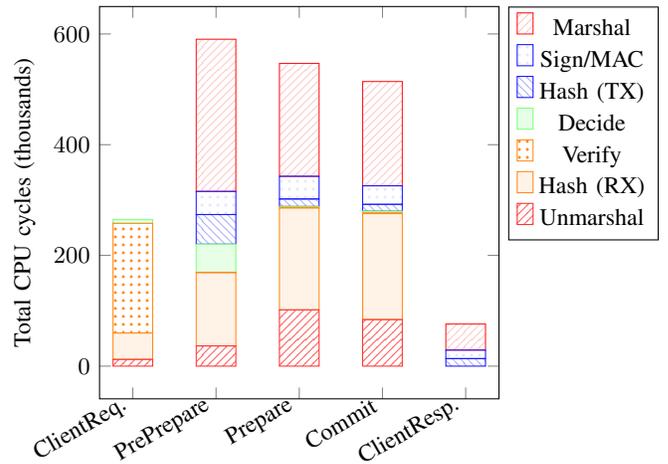
\begin{figure}[pt]
  \centering
  \begin{tikzpicture}
\pgfplotstableread{
type	pos	send	rcv	hash-sig	sig	hash-verif	verif	dec
ClientReq.	1	0	12.16831	0	0	47.58666	198.16192	6.71297
PrePrepare	2	274.8787067	36.56001067	53.43954	41.824552	132.1006867	0.141941287	51.55898
Prepare	3	203.94682	101.6559133	13.489498	40.94942133	183.8724067	1.305355333	1.6261696
Commit	4	188.3574	84.084	12.172094	33.55326333	191.3583467	1.496720867	3.1152
ClientResp.	5	47.05860133	0	13.597848	15.4	0	0	0
}\datatable

\begin{axis}[
  legend style={at={(1.4,1)},anchor=north east},
  font=\small,
  width=0.38\textwidth,
  height=0.375\textwidth,
   ybar stacked,
  bar width=15pt,  
  legend reversed,  
  ylabel={Total CPU cycles (thousands)},
    xticklabels from table={\datatable}{type},
    xtick=data,
    x tick label style={rotate=30,anchor=east},  
    ]

\addplot [red!80, pattern=north east lines, pattern color=red!60] table [x expr=\coordindex, y=rcv] {\datatable};
\addplot [orange!90, fill=orange!10] table [x expr=\coordindex, y=hash-verif] {\datatable};
\addplot [orange!90, pattern=dots, pattern color=orange!90] table [x expr=\coordindex, y=verif] {\datatable};
\addplot [green!50, fill=green!10] table [x expr=\coordindex, y=dec] {\datatable};
\addplot [blue!90, pattern=north west lines, pattern color=blue!30] table [x expr=\coordindex, y=hash-sig] {\datatable};
\addplot [blue!90, pattern=dots, pattern color=blue!10] table [x expr=\coordindex, y=sig] {\datatable};
\addplot [red!90, pattern=north east lines, pattern color=red!20] table [x expr=\coordindex, y=send] {\datatable};

\legend{Unmarshal, Hash (RX), Verify, Decide, Hash (TX), Sign/MAC, Marshal}

\end{axis}
\end{tikzpicture}

  \caption{\label{fig:cost-breakdown} The aggregate cost of each pipeline stage in a domain-optimized setup (in CPU cycles) at the leader for 15 nodes, processing 4KB values, shows that (un)marshaling and hashing before verification are the most expensive.}
\end{figure}

\subsection{What operations are costly beyond signatures?}
To understand the reason why crypto acceleration provides diminishing results in the domain-optimized case, we show a breakdown of compute costs for each message type in the leader of a group of 15 in Figure~\ref{fig:cost-breakdown} (we ignore parallelism here and compute the aggregate time spent on each part). Thanks to the MAC optimization, the biggest cost in handling requests is the time it takes to (un)marshal them and to compute their SHA256 hash for verification. With larger consensus groups, the relative cost of these operations will increase further as there will be more messages sent between nodes. When using smaller values, the cost of hashing is reduced linearly, but the cost of (un)marshaling is not reduced significantly. Hence, for simplicity we omit other data sizes from this discussion. 
If we compare these costs with an implementation that signs client responses with RSA2048, this would increase the Signing cost to the order of 4.5m cycles (around 1.2\,ms on our CPUs). This illustrates why avoiding its computation on client responses lead to speedup in Figure~\ref{fig:tput-variants} (PK-Only vs. MACs).


\subsection{Additional Evaluation of Our Prototype}

In this subsection we look at our prototype implementation as a domain-optimized BFT consensus service with reconfiguration and measure its performance in a cluster to show that it achieves low latency and high throughput even without hardware acceleration, making it a realistic starting point for implementing such functionality. 

\begin{figure}[pt]
  \centering
\begin{tikzpicture}  
  \pgfplotstableread{ 
nodes   tput512 bw512 tput4096  bw4096  tput8192  bw8192
3 47.271 0.816258449 23.716 1.707235504 16.510 2.220969372
5 36.342 1.233441572 18.201 2.60962068  14.233 3.820847644
7 29.228 1.479288655 16.825 3.613489985 10.882 4.378668201
9 26.772 1.80133365  14.745 4.219433797 8.550  4.585402575
11  22.141 1.858879537 11.756 4.203375795 7.103  4.7606553
13  18.800 1.891818447 11.777 5.051658764 6.430  5.170741561
15  16.155 1.894995097 10.205 5.105908151 5.623  5.274858219
  }\datatable
  \begin{axis}[
   legend style={at={(0.9,1.05)},
    anchor=north},
  legend columns=1,
  font=\small,
  width=0.4\textwidth,
  height=0.23\textwidth,
  ylabel={Throughput [Kops/s]},
  xlabel={Number of nodes}, 
  ytick={0,10.000,20.000,30.000,40.000,50.000,60.000},
  ymin=0, 
  ymax=55.000,
  xtick=data,
  grid=both,
  ]
  \addplot [blue!90, mark=*] table [x=nodes, y=tput512] {\datatable};  
  \addplot [orange!90, mark=triangle] table [x=nodes, y=tput4096] {\datatable};    
  \addplot [red!90, mark=o] table [x=nodes, y=tput8192] {\datatable};    

  \legend{V=512, V=4096, V=8192}
  \end{axis}
  \end{tikzpicture}  
  \caption{\label{fig:tput-nodes} Increasing the number of nodes has a predictable impact on throughput. The leader node becomes bound on its network stack for larger group sizes.}
\end{figure}
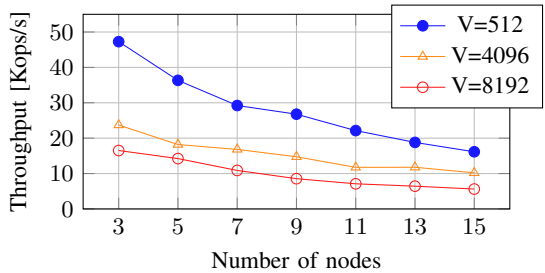

In Figure~\ref{fig:tput-nodes} we measure the throughput of our prototype with increasing consensus group sizes. The behavior is in line with the expectations of a leader-based protocol. The system delivers for 15 nodes more than 17\,kops/s for the smallest value size and 10\,kops/s respectively 5\,kops/s for the 4 and 8\,KB value sizes. The expectation is to scale to larger groups without issues, with a linear decrease in performance. 
The demonstrated throughput numbers are high enough to ensure that integration of \System  with blockchains such as Hyperledger Fabric~\cite{androulaki2018hyperledger} is possible without becoming an immediate bottleneck.

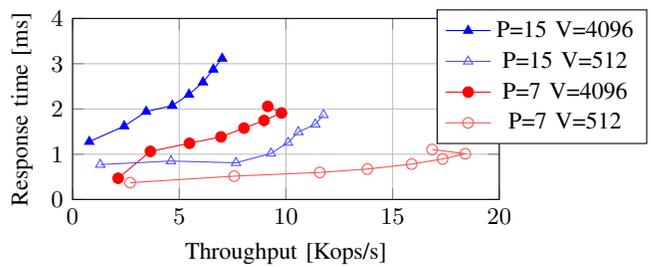
\begin{figure}[pt]
  \centering
\begin{tikzpicture}  
  \pgfplotstableread{ 
t7v512  l7v512  t15v512 l15v512 t7v4096 l7v4096 t15v4096  l15v4096
2.696  0.370919881 1.297  0.771010023 2.134  0.468603561 .783 1.277139208
7.571  0.515198351 4.612  0.849617672 3.648  1.062699256 2.420  1.620745543
11.585 0.597371565 7.659  0.809716599 5.477  1.242236025 3.449  1.941747573
13.819 0.672043011 9.300  1.018329939 6.955  1.383125864 4.673  2.074688797
15.894 0.78125 10.103 1.257861635 8.034  1.577287066 5.456  2.320185615
17.344 0.894454383 10.583 1.490312966 8.973  1.745200698 6.111  2.590673575
18.406 1.01010101  11.377 1.661129568 9.790  1.912045889 6.597  2.873563218
16.853 1.104972376 11.763 1.869158879 9.148  2.057613169 7.012  3.115264798
  }\datatable
  \begin{axis}[
   legend style={at={(1.1,1.05)},
    anchor=north},
  legend reversed,
  legend columns=1,
  font=\small,
  width=0.4\textwidth,
  height=0.22\textwidth,
  ylabel={Response time [ms]},
  xlabel={Throughput [Kops/s]},
  xmin=0,
  xmax=20.000,
  ymin=0,
  ymax=4,
  grid=both,
  ]
  \addplot [red!70, mark=o] table [x=t7v512, y=l7v512] {\datatable};  
  \addplot [red!99, mark=*] table [x=t7v4096, y=l7v4096] {\datatable};    
  \addplot [blue!70, mark=triangle] table [x=t15v512, y=l15v512] {\datatable};  
  \addplot [blue!99, mark=triangle*] table [x=t15v4096, y=l15v4096] {\datatable};  

  \legend{P=7 V=512, P=7 V=4096, P=15 V=512, P=15 V=4096}

  \end{axis}
  \end{tikzpicture}
  \caption{\label{fig:tput-lat} Even under load the prototype delivers response times that increase predictably. For small values it is possible to keep response times under 2ms even when fully saturated.}
\end{figure}

With the deployment of permissioned ledgers in datacenter-like environments, it is important that the underlying BFT consensus can be performed with low latency. As shown in Figure~\ref{fig:tput-lat}, the average response time of our prototype starts from the sub-millisecond range for small values and increases slowly with load. Even close to saturation, the response time is only factor of three larger then in the unloaded case. While it is not our focus to compare to BFT-Smart since our prototype is meant as a platform for future exploration, not as a production-ready solution, it is worth pointing out that the latency stays \emph{under} 3.5ms at all times, which is the lowest measured in Figure~\ref{fig:bft-smart}.

Overall, low and predictable latency that does not increase significantly with load is important because it ensures that the consensus nodes will not be the latency bottleneck. Even though most permissioned blockchains today do not optimize for latency at the millisecond level, as we discuss in the Related Work section, there are emerging blockchain designs, e.g.~\cite{istvan2018streamchain,amiri2019caper}, that could readily benefit from lower latency consensus.

\section{Insights and Acceleration Strategies}
\label{sec:discussion}

Related works have shown that replacing PK signatures with MACs can improve performance but the improvements are seldom quantified. In this work, by measuring the difference in the same system, we reach a counter-intuitive insight: when adding crypto acceleration to the most optimized version, the performance gains are only marginal because client signatures can be verified in parallel, and block signatures can be computed on the side of normal operation. As the cost drill-down shows, for the domain-optimized case, the more significant opportunities are in acceleration of data movement and hashing. These will provide a bigger benefit than focusing solely on crypto accelerators. In the remaining we discuss two promising acceleration strategies to help reach 10Gbps line-rate performance and beyond. In preparation for their implementation, we present micro-benchmarks further motivating them and discuss their main benefits and challenges.  


\subsection{Offloading to SmartNICs}
Figure~\ref{fig:cost-breakdown} shows that (un)marshaling and hashing costs account for a significant portion of the runtime even if we don't factor in signature verification. Today, there is an emerging offering of SmartNICs~\cite{firestone2018azure,williams2019exploring,eran2019nica} that, in the future, could be used to offload some of these operations, e.g., serialization of messages and line-rate hashing. Furthermore, there are recent related works that use Mellanox NICs to offload TLS~\cite{pismenny2016tls}, which could be used as an equivalent of MACs.

The main question that arises when proposing the use of such SmartNICs is whether to treat them \emph{a)} as a ``stateless'' accelerator that can, for instance, parse packets but does not access application state, or \emph{b)} as a ``stateful'' one that can in addition also clone and send messages to different peers depending on the application state. 

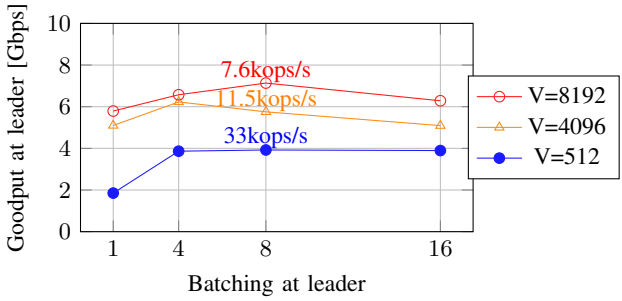
\begin{figure}[pt]
  \centering
\begin{tikzpicture}  
  \pgfplotstableread{ 
batch t5120 t4096 t8192 bw512 bw4096  bw8192
1 15787 10192 6174  1.85182839 5.09940381 5.79174367
4 32940 12450 7008  3.86388972 6.22915791 6.57410749
8 33440 11500 7606  3.92254014 5.75384064 7.13508298
16  33200 10176 6700  3.89438794 5.09139847 6.28517696
  }\datatable
  \begin{axis}[
   legend style={at={(1.18,0.75)},
    anchor=north},
  legend columns=1,
  legend reversed,
  font=\small,
  width=0.375\textwidth,
  height=0.24\textwidth,
  ylabel={Goodput at leader [Gbps]},
  xlabel={Batching at leader}, 
  ymin=0, 
  ymax=10,
  xtick=data,
  grid=both,
  ]
  \addplot [blue!90, mark=*] table [x=batch, y=bw512] {\datatable};  
  \addplot [orange!90, mark=triangle] table [x=batch, y=bw4096] {\datatable};    
  \addplot [red!90, mark=o] table [x=batch, y=bw8192] {\datatable};    

  \node[blue] at (axis cs: 8,4.5) {33kops/s};
  \node[orange] at (axis cs: 8,6.3) {11.5kops/s};
  \node[red] at (axis cs: 8,7.7) {7.6kops/s};

  \legend{V=512, V=4096, V=8192}
  \end{axis}
  \end{tikzpicture}  

  \caption{\label{fig:tput-bw} Our prototype takes advantage of the 10Gbps bandwidth available at the leader, but to achieve an almost complete utilization of it some amount of batching is necessary.}
\end{figure}

The first case already allows one to offload hashing and parsing, as well as, marshaling to the device, but to estimate the benefits of exposing more of the protocol's state to the NIC, we need to evaluate how efficient the current software version is when interfacing with the network. To this end, we plot in Figure~\ref{fig:tput-bw} the useful bandwidth usage at the leader (goodput) with an increasing batching factor. Batching is implemented in the leader node by waiting for multiple messages from clients, assembling them into a vector and issuing a single $\PREPREPARE$ message for them. If there would be no TCP/IP and Ethernet overhead, it would be possible to achieve at most 10Gbps goodput in our setup, but in reality, even for very large packets the limit is lower. The results show that without batching, it reaches up to 6Gbps goodput for large values (4 and 8\,KB) and around 2Gbps for small ones (512\,B). Moderate batching of 4 to 8 requests can result in a better TCP stack utilization and at the leader goodput can reach more than 7Gbps for large requests and 4Gbps for small ones. This comes at the cost of higher response times, though with these batching factors, the differences remain small.

Based on this result, we foresee that for network speeds beyond 10Gbps, SmartNICs will be a sensible acceleration option. They will have to offload parts of the packet-processing operations and rely on fine-grained batching with strict latency guarantees that would be unfeasible to achieve in software.

\subsection{PBFT on Standalone FPGAs}
\label{sec:standalone}
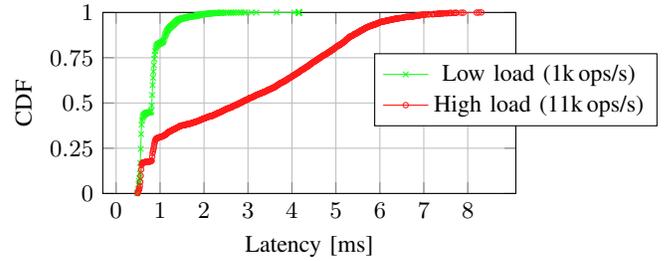
\begin{figure}[pt]

  \centering
  \begin{tikzpicture}  
  \begin{axis}[
   legend style={at={(1,0.775)},
    anchor=north},
  legend columns=1,
  font=\small,
  width=0.39\textwidth,
  height=0.22\textwidth,
  ylabel={CDF},
  xlabel={Latency [ms]}, 
  xtick={0,1,2,3,4,5,6,7,8},
  ytick={0,0.25,0.5,0.75,1},
  ymin=0, 
  ymax=1,
  grid=both,
  ]
  \addplot [green!90, mark=x, mark options={scale=0.75}] table [x expr={0.001*\thisrowno{2}}, y=pct0] {latency-cdf.dat};  
  \addplot [red!90, mark=o, mark options={scale=0.5}] table [x expr={0.001*\thisrowno{0}}, y=pct1] {latency-cdf.dat};  
  
  \legend{Low load (1k\,ops/s), High load (11k\,ops/s)}
  \end{axis}
  \end{tikzpicture}  

  \caption{\label{fig:jitter} Our prototype implementation highlights additional opportunities for HW: reducing the long tail of response times under load (15 nodes and 4KB payload)}
\end{figure}

Various types of hardware accelerators have been recently used to accelerate CFT consensus~\cite{istvan2016consensus,dang2018p4xos,wang2017apus,poke2015dare,jin2018netchain}. 
These solutions demonstrate latencies in the order of microseconds and are able to saturate the network regardless of the value sizes, thanks to the reduced overhead of the network stack and the low cost of data movement between network interface and the decision logic. They also bring predictable response time behavior which enables them to fulfill strict SLAs in low latency environments. If we investigate the distribution of response times and the variance at the tail of our software prototype, we see a significant increase in the high percentiles of response times, even if the median does not shift by much (Figure~\ref{fig:jitter}). In deployments where SLAs are important, standalone FPGA-based implementations could be beneficial because there are less factors influencing response times. 

Even though hardware-based solutions provide microsecond latencies and high throughput, one challenge in this space has been the feasibility of handling not only the failure-free case but implementing reconfiguration as well. There is prior work, e.g.~\cite{istvan2016consensus}, that demonstrates that it is possible to implement reconfiguration for an atomic broadcast protocol on FPGAs and, in terms of communication patterns and metadata structures, PBFT is not significantly more complex. However, there is a significant difference between BFT and CFT algorithms in that the former requires the computation and verification of cryptographic hashes and signatures. This additional requirement, in particular RSA, make implementation challenging on FPGAs. This is because RSA (and similar ciphers) require iterative computation that is, on the one hand, resource intensive and, on the other hand, suffers from the relatively low clock rates of FPGAs. Therefore, from the three BFT variants discussed in this paper, only the domain-optimized is feasible on FPGAs because it minimizes the need for cryptographic operations, i.e., the rate of RSA ops/s. 

\begin{figure}[pt]
  \centering
\begin{tikzpicture}  
  \pgfplotstableread{ 
GroupSize	Use512	Use4096	rsa512	rsa4096
7		247		104		190		50
16		147		74		90		20
25		107		74		50		20
34		97		64		40		10
43		87		64		30		10
52		87		64		30		10
  }\datatable
  \begin{axis}[
   legend style={at={(1.35,1.15)},
    anchor=north},
  legend columns=1,
  font=\small,
  width=0.32\textwidth,
  height=0.22\textwidth,
  ylabel={Logic slices (thousands)},
  xlabel={Smallest targeted group size}, 
  ymin=0, 
  ymax=250,
  xmin=7,
  xmax=43,
  xtick=data,
  grid=both,
  ]

  \draw[color=gray, line width=2pt] (axis cs:7,108) -- (axis cs:52,108);
  \node[color=gray] at (axis cs: 33,130) {xc7vx690t};

  \addplot [blue, mark=*] table [x=GroupSize, y=Use512] {\datatable};  
  \addplot [blue, dashed] table [x=GroupSize, y=rsa512] {\datatable};

  \addplot [orange, mark=triangle] table [x=GroupSize, y=Use4096] {\datatable};    
  \addplot [orange, dashed] table [x=GroupSize, y=rsa4096] {\datatable};

  \legend{V=512 All logic, V=512 Just RSA, V=4096 All logic, V=4096 Just RSA}
  \end{axis}
  \end{tikzpicture}  
  \caption{\label{fig:fpga-mockup} Estimated resource consumption of a 10Gbps PBFT implementation on FPGAs using open-source components targeting a minimum value size and group size.}
\end{figure}
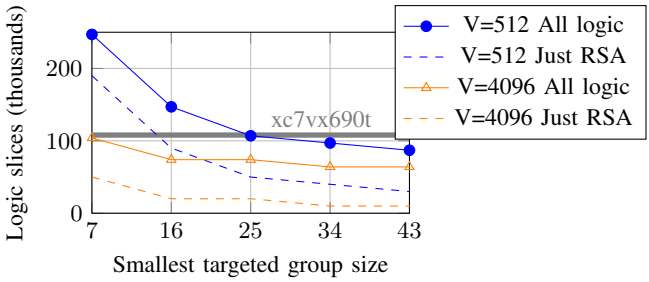

To verify this claim, we rely on open source cores to estimate the cost of an implementation. By computing the maximum 10Gbps client-facing throughput at the leader as a function of minimum consensus group size and minimum value size, we can estimate how many resources RSA-related computations would take up on the FPGA. We use the RSA core from the Xilinx Vitis Library~\cite{repoVitis} as a representative instance and replicate it as many times as needed to match the desired throughput level. 
In Figure~\ref{fig:fpga-mockup} we show with dashed lines how the cost of RSA computations decreases as the minimum group size increases. This is because the leader will send increasingly more intra-node messages than to/from clients. 

To estimate the total cost of a complete 10Gbps BFT implementation in terms of logic resources, we synthesized a 10Gbps TCP/IP module with DRAM controller~\cite{repoTCP}, SHA256 hashing cores~\cite{repoSHA}, and AES~\cite{repoVitis} cores. For an estimate of the decision logic we relied on the CFT Atomic Broadcast module in Caribou~\cite{repoCaribou}. The resource cost of these modules was added to the cost of RSA cores (sum shown as solid lines).

To put the logic resource numbers in perspective, we show the capacity of a mid-range FPGA\footnote{Xilinx xc7vx690t: 108k Logic Slices, 3600 DSPs and 1470 BRAMs.} with the horizontal line. Overall, this resource estimation is showing an encouraging result: when using larger value sizes, or larger groups, even a mid-range FPGA could implement PBFT at 10Gpbs line-rate. Furthermore, given that there are also larger FPGAs on the market, e.g., Amazon F1 instances have around 4 times larger FPGAs, we are confident that it is realistic to implement in the future BFT on stand-alone nodes.

\section{Related Work}

\subsection{Protocol Variants and Optimizations}

There is related work on implementing traditional BFT protocols in such a way that benefits from the multi-core parallelism of modern CPUs. A representative example is BFT-Smart~\cite{bessani2014state}. The pipelined implementation demonstrates good performance on Gigabit networks. The way in which reconfiguration is performed is similar to the one we propose in \System  but the nodes have to do more complex operations to perform state transfer. We chose not to use BFT-Smart as the framework for this study because it has implicit design choices related to, for instance, signatures, that would have made it difficult to ``simulate'' different BFT variants. The results of this work, nonetheless, apply to systems such as BFT-Smart.

There is an increasing interest in adopting the PBFT protocol for permissioned blockchain purposes. SBFT~\cite{gueta2019sbft}, for instance, aims to reduce communication complexity by relying on an so called ``collector'' node and threshold signatures, and adding a fast path to the execution (similar to Zyzzyva~\cite{kotla2007zyzzyva}). SBFT targets wide area networks and trades off bandwidth for more compute-intensive operations. As we shown in this work, however, in fast networking deployments there is plenty of bandwidth and relying on private-key cryptography as the default leads to sub-optimal use of the network resources.

Other recent work, such as HotStuff~\cite{yin2019hotstuff}, explores how view-changes can be made cheaper. It targets permissioned blockchains that experience a high amount of failures or churn among the consensus nodes and, as a result, will require frequent view changes. The authors reduce the cost of these operations by adding an extra communication phase to each consensus round. In this work we assume business-to-business use-case of permissioned ledgers where, even though the clients of the system can be subject to churn, the core consensus nodes rarely change. In this setting optimizing for failure-free behavior is more beneficial. Nonetheless, our findings will apply to solutions such as HotStuff, as long as they are being executed in low latency environments.

Not surprisingly, the performance of PBFT and any similar protocol, including \System, is severely limited by the leader as consensus groups grow. There is emerging work~\cite{stathakopoulou2019mir} that aims to solve the leader bottleneck without fundamentally changing the underlying protocol and instead relying on deterministic scheduling and data sharding that fits the permissioned ledger use-case well. Mir-BFT~\cite{stathakopoulou2019mir} achieves a near-linear increase in throughput with each node added to the consensus group and is competitive even when compared to ring replication in terms of bandwidth usage. The findings of this work are directly applicable to Mir, since its multi-leader approach is fully orthogonal to the actual implementation of the BFT protocol underneath.

\subsection{Consensus and Specialized Hardware}

Various types of hardware accelerators have been used to accelerate CFT consensus algorithms~\cite{istvan2016consensus,dang2018p4xos,wang2017apus,poke2015dare} and they demonstrate latencies in the tens of microseconds and are able to saturate the network regardless of the value sizes, thanks to reducing the overhead of the network stack and the data movement between network interface and the decision logic. We believe that there is an emerging opportunity in exploring how these ideas can translate to BFT consensus.

Other related work uses specialized hardware to implement trusted computing elements and through this simplify the typical three-round operation of BFT to two rounds~\cite{behl2017hybrids,kapitza2012cheapbft} and reduce the number of necessary replicas to $2f+1$. Even thought they show promising result and are well suited to fast networks, these works introduce a different trust model for the two ``parts'' of the nodes.

\section{Conclusion}

In this work we deconstructed a BFT consensus protocol with the goal of forecasting the benefits of various acceleration strategies. Our work is motivated by the emergence of permissioned blockchain use-cases that can be ran in environments with high bandwidth networking and low latencies and should be able, in the future, to take advantage of a wide range of acceleration options. Based on our study, comparing different BFT consensus variants, we concluded that the key to achieving low latency and high throughput behavior is more complex than just offloading cryptographic operations and instead will require a clever combination of improvements to multiple steps of the processing pipeline. This finding is a catalyst for research into hybrid solutions, that combine software and hardware in surprising ways.

\section*{Acknowledgments}
This project has received funding from the European Union’s
Horizon 2020 research and innovation program, under the
Marie Sklodowska-Curie grant agreement No. 842956, and
the Spanish Research Council, through the Juan de la Cierva
Formacion funding scheme.

\bibliographystyle{abbrv}
\bibliography{biblio}


\end{document}